\documentstyle[prd,aps,preprint]{revtex}
\begin{document}
\draft

%
%  Uncomment following two lines and one below for 2 column format.
%
%\twocolumn[\hsize\textwidth\columnwidth\hsize\csname
%@twocolumnfalse\endcsname

\preprint{Nisho-00/1} \title{Ultra High Energy Cosmic Rays \\
from Axion Stars } 
\author{Aiichi Iwazaki}
\address{Department of Physics, Nishogakusha University, Chiba
  277-8585,\ Japan.} \date{Mar 1, 2000} \maketitle
\begin{abstract}
We propose a model in which ultra high energy cosmic rays are produced by 
collisions between neutron stars and axion stars.
The acceleration of such a cosmic ray is made by the electric field, 
$\sim 10^{15}\,\,(B/10^{12}\,\mbox{G})\,\,\mbox{eV}\,\,\mbox{cm}^{-1}$, 
which is induced in 
an axion star by the strong magnetic field $B>10^{12}$ G of a 
neutron star. 
As we have shown previously,
similar collisions generate gamma ray bursts when the magnetic field
is much smaller, e.g. $\sim 10^{10}$ G. 
If we assume that
the axion mass is $\sim 10^{-9}$ eV, 
we can explain huge energies 
of the gamma ray bursts $\sim 10^{54}$ erg as well as 
the ultra high energies of the cosmic rays $\sim 10^{20}$ eV.
In addition, it turns out that
these axion stars are plausible candidates for MACHOs.
We point out a possibility of observing monochromatic radiations
emitted from the axion stars.

\end{abstract}
%\pacs{73.61.-r,73.20.Dx,73.40.Hm,73.40.Gk}

\vskip .7cm

\pacs{14.80.Mz, 95.30.+d, 98.80.Cq, 98.70.Rz, 97.60.Jd, 05.30.Jp, 98.70.Sa, 
11.27.+d 
%\\Axion, Dark Matter, Ultra High Energy Cosmic Ray, Gamma Ray Burst, 
%Boson Star 
\hspace*{3cm}}
\vskip2pc
%%%%%%%%%%%%%%%%%%%%%%%%%%%%%%%%%%%%%%%%%%%%%%%%
\tightenlines
The origin of ultra high energy cosmic rays ( UHECRs ) is 
one of most mysterious puzzles in astrophysics\cite{uhe}.
UHECRs with energies $\sim 10^{20}$ eV can not travel in distance 
more than $50$ Mpc 
owing to the interactions between UHECRs and the cosmic backgrand radiations.
Observations, however, show that there are no visible candidates for 
the generators of such UHECRs in the arrival directions of UHECRs.  
On the other hand, dark matter in the Universe\cite{text} is
one of most mysterious puzzles in cosmology. 
Axion\cite{PQ,kim} is a plausible candidate for the dark matter.  
Probably, some of axions may form boson stars ( axion stars ) 
in the present Universe
by gravitational cooling\cite{cooling} or 
gravitational collapse of axion clumps formed at the period of 
QCD phase transition\cite{kolb}. Here we explain a generation mechanism
of UHECRs and discuss a production rate of UHECRs 
assuming collisions between 
axions stars and neutron stars. We also show that the collisions 
cause emission of observable monochromatic radiations from the axion stars.

We have previously proposed\cite{iwa} a possible generation mechanism 
of gamma ray burst ( GRB ).
According to the mechanism 
the collision between  
axion star and neutron star 
generates a gamma ray burst; the axion star 
dissipates
its mass energy\cite{iwa,iwa1} very rapidly under the strong magnetic field 
of the neutron star. Thus, the energy released in the collision is given by 
the mass $M_a$
of the axion star. Typically it is given by 
$M_a\sim 10^{-5}M_{\odot}\,(10^{-5}\mbox{eV}/m_a)$ where
$m_a$ denotes the axion mass. 
In the analysis we have taken 
the mass such as $m_a\sim 10^{-5}$ eV as suggested observationally
in standard invisible axion models.  The mass, $M_a\sim 10^{49}$ erg,
corresponding to this choice, however,   
is not enough to explain a huge energy observed in GRB980123
even if jet is assumed in the GRB.

In this paper, we explain an acceleration mechanism of UHECRs
assuming the axion mass, $m_a\sim 10^{-9}$ eV,
although the choice is not conventional\cite{text,string}. The essence of 
the acceleration is that  
a strong electric field 
$\sim 10^{15}\,\,(B/10^{12}\mbox{G})\,\,\mbox{eV}\,\,\mbox{cm}^{-1}$
is induced in the axion star when it is exposed to the magnetic field
$B>10^{12}$ G of a neutron star. This electric field can accelerate
charged particles and makes them obtain the huge energies $\sim 10^{20}$ eV. 
Furthermore, it turns out that   
the energy $\sim 10^{54}$ erg observed in the GRB
can be also explained assuming a moderate jet of the GRB.
This is because the mass
of the axion star is given by $\sim 10^{-1}M_{\odot}\simeq 10^{53}$ erg
with the assumption of $m_a$.
%that charged particles are accelerated to gain high energies 
%such as $10^{20}$ eV by the strong electric field induced
%inside of the axion star when it is under the strong magnetic field 
%$> 10^{12}$ G. 
Thus the collisions 
between the axion stars and the neutron stars are possible sources of 
both UHECRs and GRBs. 
Additionally, it turns out that the axion 
star is a plausible candidate for MACHO\cite{MACHO} 
because the value of the mass 
is suitable for explaining the
observations of MACHOs. Since all of baryonic
candidates for MACHOs have been argued to have serious difficulties\cite{non},
nonbaryonic ones like the axion stars are favored. 
We discuss that 
the collisions generate monochromatic radiations
with a frequency $m_a/2\pi\simeq 2.4\times 10^5$ Hz. With the detection of
such radiations we can test our model and determine the axion mass.

%As far as we know, 
%this axion star is only such a 
%candiadte for explaining not only MACHOs but also
%ultra high energy cosmic rays and gamma ray bursts.

As we shall show below, we need to separate two cases of 
the collisions, the collision with a
neutron star possessing relatively weak magnetic field $\sim 10^{10}$ G and 
the one with a neutron star possessing 
relatively strong magnetic field $ > 10^{12}$ G.   
In the first case the axion star collides directly with the neutron star
and dissipates its whole energy inside of the outercrust
of the neutron star\cite{iwa}.
The ultra high energy cosmic rays are not produced in this case but 
gamma ray bursts are produced.
On the other hand, in the second case the axion star never collide 
directly with the neutron star because it evapolates before the collision.
This is because the axion star induces a stronger electric field\cite{iwa,iw} 
when it is exposed to the stronger magnetic field of the neutron star. 
But such a strong electric field decays very rapidly owing to 
electron-positron pair creations\cite{vacdecay} 
when the strength of the electric field goes beyond
a critical value. 
Namely, as the axion star approaches the neutron star,
the strength of the magnetic field around it increases gradually and 
reaches a critical value beyond which the electric field induced 
in the axion star decays very rapidly. It means that the axion star
itself decays very rapidly.
In this case the ultra high energy 
cosmic rays are produced.
Gamma ray bursts might be also produced, but they are emitted  
in a cone with much small solid angle and 
their duration is very short ( less than millisecond ).

Let us first explain briefly axion stars ( ASs ) 
and how they induce strong electric fields  
under a magnetic field of a neutron star.
The AS is a coherent object of the real scalar field 
$a(x)$ describing the axion.
It is represented by a solution\cite{real,iwasol}
of the equation of the axion field coupled with gravity.
An approximate form of the solution\cite{iwasol} is given by

\begin{equation}
\label{a}
a(x)=f_{PQ}a_0\sin(m_at)\exp(-r/R_a)\,, 
\end{equation}
where $t$ ( $r$ ) is time ( radial ) coordinate and 
$f_{PQ}$ is the decay constant of the axion. 
The value of $f_{PQ}$ is constrained\cite{text} conventinally 
from cosmological 
and astrophysical considerations\cite{text,kim} such that 
$10^{10}$ GeV $< f_{PQ} <$ $10^{12}$ GeV ( the axion mass $m_a$ is 
given in terms of $f_{PQ}$ such that $m_a\sim 10^7\,\,\mbox{GeV}^2/f_{PQ}$ ).
However, when we assume unconventionally entropy productions 
below the temperature $1$ GeV in the early Universe, we may be 
released from the constraints\cite{entropy}.
Hereafter we assume that $f_{PQ}\sim 10^{16}$ GeV or $m_a\sim 10^{-9}$ eV.

In the formula,
$R_a$ represents the radius of 
an AS which has been obtained\cite{iwa1,iwasol} numerically 
in terms of mass $M_a$ of the AS;
$R_a=6.4\,m_{pl}^2/m_a^2M_a\simeq 1.6\times 10^5\,\mbox{cm}\,
m_9^{-2}\,(10^{-1}M_{\odot}/M_a)$,
with $m_9=m_a/10^{-9}$ eV and $m_{pl}$ is Planck mass.
Similarly, the amplitude $a_0$ in eq(\ref{a}) is represented such as  
$a_0=1.73\times 10^2\,(10^{5}\,\mbox{cm}/R_a)^2\,
m_9^{-1}$.
Therefore, 
we find that the solution is parameterized by one free parameter,
either one of the mass $M_a$ or the radius $R_a$.
It is also important to note that the solution is not static but
oscillating with the frequency of $m_a/2\pi$. 
It has been demonstrated\cite{re} 
that there is no static 
regular solution of the real scalar massless field coupled with gravity.
This may be general
property of the real scalar massive field. 
%On the other hand, 
%static solutions\cite{re}
%exist in the case of the complex scalar field. 

The AS mass is determined by physical conditions under which the AS
has been formed; how large amount of cloud of axions are cooled gravitationally
to form the AS, etc.. The situation is similar to other stars such as
neutron stars or white dwarfs. A typical mass scale in these
cases is the critical mass\cite{star}; 
stars with masses larger than the critical 
mass collapse gravitationally into more compact ones or black holes. 
In the case of the AS, there also exists a critical mass $M_c$ 
which is given by\cite{real}
$M_c\simeq 10^{-1}M_{\odot}\,m_9^{-1}$ 
where $M_{\odot}$ represents the solar mass.
Therefore, we adopt this critical mass $M_c$ 
as a typical mass scale of the ASs. 
%Thus, the corresponding radius $R_a$ of the ASs 
%with this critical mass
%is given such that $R_a\simeq 1.6\times 10^{5}\,m_9^{-1}\,\mbox{cm}$. 
( The critical mass is the maximal mass which ASs can take. Thus, 
their masses, in general, are smaller than this one. Since energies released
in GRBs by the ASs are given by masses themselves, the maximal energy  
in GRBs is given by the critical mass ).

Although the gravitational cooling for star formation is in general 
ineffective because it is too slow process, it has been shown\cite{cooling} 
that the cooling is very effective for the real scalar axion field.
Thus, the axion stars can be 
easily formed gravitationally in a gas of the axions.
It is reasonable to assume that the most of the axions in the Universe
forms the axion stars.

Let us explain how an AS induces an electric field 
under a magnetic field $\vec{B}$ of a neutron star. Owing to the interaction 
between the axion and the electromagnetic field
described by
$ L_{a\gamma\gamma}=c\alpha a(x)\vec{E}\cdot\vec{B}/f_{PQ}\pi$, 
where the value of $c$ ( of the order of unity )
depends on axion models\cite{DFSZ,hadron,kim},
the Gauss law is modified\cite{Si} such that 
$\vec{\partial}\vec{E}=-c\alpha \vec{\partial}\cdot(a(x)\vec{B})/f_{PQ}\pi
+\mbox{``matter''}$.
The last term ``matter'' denotes electric charges of ordinary matters.
The first term in the right hand side 
represents an electric charge made of the axion. ( This charge is oscillating
and so there exist a corresponding oscillating current, 
$J_a=c\,\alpha\,\partial_ta(x)\vec{B}/f_{PQ}\pi$.
Thus, radiations are emitted by the AS, which might be observable. 
We will discuss it later. )
Thus, we find that  
the electric field $\vec{E_a}$ is induced;  
$\vec{E_a}=-c\,\alpha \,a(x)\,\vec{B}/f_{PQ}\pi$ with $\alpha\simeq 1/137$, 
Numerically, its strength is given by 

\begin{equation}
\label{e}
E_a \sim 10^{15}\,\,\mbox{eV}\,\,\mbox{cm}^{-1}\,\,B_{12}\,m_9\, 
\quad ,
\end{equation}
with $B_{12}=B/10^{12}$ G,
where we have used the solution in eq(\ref{a}) for the critical mass. 
Obviously, the spatial extention of the field is given by 
the radius $R_a\simeq 1.6\times 10^{5}\,m_9^{-1}\,\mbox{cm}$ of the AS.

This electric field is oscillating with the frequency, 
$m_a/2\pi\simeq 2.4\times 10^5\,m_9\,\mbox{Hz}$.
Thus a particle with electric charge Ze can be accelerated in a direction 
within the half of the period,
 $\pi/m_a$ or
in a distance $\simeq R_a$ 
( $\sim \pi/m_a\times\mbox{light velocity}$ ) by this field, 
unless it collides with other particles within the period.
Thus, the energy $\Delta E$ obtained by the particle is given by

\begin{equation}
\label{delta}
\Delta E=\mbox{Ze}\,E_a\times \pi/m_a
\times \mbox{light velocity}\sim 10^{20}\,\mbox{Z}\,\,\mbox{eV} B_{12}\quad .
\end{equation}

Therefore, the electric field  
can accelerate the charged particle so that its energy
reaches $\sim 10^{20}\,\mbox{Z}$ eV. 
These charged particles may be 
baryons or electron-positron pairs produced 
by the decay of the electric field itself as discussed soon below.

Comment is in order. It seems apparently from eq(\ref{delta})
that stronger magnetic fields 
yield cosmic rays with higher energies. 
Stronger magnetic fields, however, induce stronger electric fields, which
are unstable against electron-positron pair creations. Therefore, strong
magnetic fields do not necessarily yield cosmic rays with higher energies.
%There is a critical strength of the electric field below which the field is 
%stable and can accelerate charged particles. Since the critical strength 
%depends on the factor of $B_{12}\times m_9$ as is shown below, 
%and the energy $\Delta E$ of cosmic rays depends only on the factor of
%$B_{12}$, the energy $\Delta E$ can increase with the magnetic field 
%as far as $B_{12}\times m_9$ remains constant. Thus smaller mass of the axion
%can leads to cosmic rays with higher energies when the
%neutron star possesses stronger magnetic field.

If the particles collide with other particles on the way of acceleration,
in other words, their mean free paths are shorter than $R_a$,
they can not obtain such high energies. It is easy to see, however, that 
the mean free paths of quarks or leptons with much higher energies 
than their masses are longer than 
$R_a$ in magnetosphere of neutron star.
This is because since  
cross sections, $\sigma$, of quarks or leptons with such energies $E$
behaves such as $\sigma \sim 1/E^2$, 
mean free paths $\sim 1/n\sigma$ is longer than
$R_a\sim 10^5$ cm unless number density $n$ 
of particles around the AB is extremely large ( i.e.
$n>10^{40}/\mbox{cm}^3$ ).

As is well known, the strong electric field is unstable\cite{vacdecay} against 
electron-positron pair creations. 
This implies that AS decays into the pairs when the electric field 
induced in AS is sufficiently strong.

Let us estimate the decay rate of the field
and show that the AS decays before colliding with a neutron star 
whose magnetic field at surface is stronger than $10^{12}$ G.
We also show that  
the AS can collide directly with a neutron star 
whose magnetic field is relatively weak
$\sim 10^{10}$ G.

The decay rate $R_d$ of the field per unit volume and 
per unit time is given by\cite{vacdecay}

\begin{equation}
\label{Rd}
R_d=\frac{\alpha E_a^2}{\pi^2}\sum_{n=1}^{\infty}
\frac{\exp{(-m_e^2\pi n/eE_a)}}{n^2}
\end{equation}
where   
$m_e$ denotes electron mass. The rate is very small for 
an electric field much weaker than $m_e^2\pi\sim 4\times 10^{16}$ eV/cm.  
The electric field of AS, however, can be very strong 
and it can be comparable to $m_e^2\pi$. Therefore,
the rate is much large. 
Numerically, it is given by 
$R_d\sim 10^{47}\,B_{12}^2\,m_9^2\,\, \mbox{cm}^{-3}\mbox{s}^{-1}\,
\sum_{n=1}^{\infty}\exp(-0.7\times10^{2}\,n/B_{12}\,m_9)/n^2$.
Since the spatial extention of the field is approximately given by 
$10^5\,m_9^{-1}$ cm, the total decay rate $W$ of the field in AS is 
$\sim  10^{62}\,B_{12}^2\,m_9^{-1}\mbox{s}^{-1}\,\,
\sum_{n=1}^{\infty}\exp(-0.7\times 10^{2}\,n/B_{12}\,m_9)/n^2$. 
Numerically, it reads

\begin{equation}
\label{W}
W\simeq  10/\mbox{s}\,\,\mbox{for}\,\, B_{12}=0.5,
\quad W\simeq 10^6/\mbox{s}\,\,\mbox{for}\,\,B_{12}=0.55,
\quad W\simeq  10^{32}/\mbox{s}\,\,\mbox{for}\,\,B_{12}=1
\end{equation}
with $m_9=1$.

Therefore, we find that the AS decays very rapidly ( or almost suddenly )
when it approaches a region 
where the strength of the magnetic field reaches a critical value of 
about $10^{12}$ G. Hence, the AS evapolates before colliding with the neutron 
star whose magnetic field at the surface is stronger than
$10^{12}$ G.  
The whole energy of the AS is transmitted to 
the charged particles, each of which can obtain energies $\sim 10^{20}$ eV. 
These particles are emitted into a cone with
very small solid angle. They form 
an extremely short pulse whose width being less
than millisecond. Actually, when we suppose 
that the relative velocity of the AS is equal to light velocity,
it decays approximately within a period of 
$10^{-4}\mbox{sec}\sim \,\, 10^{-5}$sec; 
in the period it passes the region where
the magnetic field increases from 
$0.5\times 10^{12}$ G to $10^{12}$ G.  
These leptons may be converted into baryons and photons through 
the interactions with themselves, interstellar
medium or ejection of progenitor of the neutron star.
Thus when the magnetic field is sufficiently strong, 
ultra high energy cosmic rays can be produced.

We comment that the velocity of AS trapped gravitationally to
a neutron star, is 
approximately given by the light velocity 
just when it collides with the neutron star.
This is because an AS can be trapped to a neutron star 
when the AS approaches it within a distance $\sim 10^{11}$ cm 
as we will show below, and then, 
the AS collides with the neutron star
losing its potential energy and angular momentum
by emitting gravitational waves. Thus, the velocity of the AS
reaches approximately the light velocity 
when it collides with the neutron star.

Furthermore, we can see from eqs(\ref{e}) and (\ref{Rd}) 
that the critical electric field depends on the factor of $B_{12}\times m_9$.
On the other hand, the energy $\Delta E$ obtained by 
accelerated charged particles depends only on the factor of $B_{12}$.
Therefore, we find that if $m_a$ is smaller than 
$10^{-9}$ eV, the maximal energy of cosmic rays can be 
larger than $10^{20}$ eV when $B>10^{12}$ G.

It is easy to see that the decay rate is negligibly small for the case of 
weak magnetic field $\sim 10^{10}$ G. Thus the AS collides directly with 
such a neutron star and dissipates the whole energy 
in the outercrust of the neutron star.
Actually, the rate of the energy dissipation in the crust
has been estimated previously\cite{iwa,iwa1}
and given approximately by $10^{46}\mbox{erg/s}\,\mbox{cm}^3$, while
the energy density of the AS is given by $10^{38}\mbox{erg}/\mbox{cm}^3$.
Thus even if the velocity of the AS is equal to the light velocity,
it dissipates its whole energy in the crust: it never reaches 
the core of the neutron star.
This estimation has been performed by noting that 
the energy dissipation of the AS arises due to 
the dissipation  
of an electric current ( $=\sigma_c \,E_a$ ) induced
in the crust with conductivity $\sigma_c$;
the value of $10^{26}$/s
has been used for $\sigma_a$\cite{con}.
 
Anyway, this collision 
generates gamma ray bursts with energies $\sim 10^{53}$ erg. 
The ejection could be emitted
into a cone with small solid angle as a jet. This is because particles 
( mainly irons ) of the neutron star are accelerated and emitted
in the direction of the strong electric field $\sim 10^{13}\,B_{10}$ eV/cm. 
The fact that the whole energy of AB is dissipated only in the 
outercrust, implies that the ejection may be particles composing the 
crust. Thus a fraction of the baryon contamination in the jet is less than
$10^{-5}M_{\odot}$ as required observationally.

In the above case, the AS dissipates its whole energy  
in the first collision. On the other hand, an AS may collide several times
with a neutron star when its mass is much smaller than the critical mass
$\sim 10^{-1}M_{\odot}$. The mass has been assumed as a typical mass of ABs.
We see from the general formula of $R_a$ that the radius of the AS 
with smaller mass than 
the typical one is larger than the typical radius $\sim 10^5$ cm. 
For example, if its mass is given by $10^{-2}M_{\odot}$, the radius is about
$10^{6}$ cm. This is comparable to the radius of the neutron star.
In addition, an electric field induced in such an AS is weaker than the 
typical one eq(\ref{e}). This implies that the rate of the energy dissipation
in a neutron star is smaller than the one quoted above.
In such a case the collisions might occur several times. We expect that 
these collisions leads to GRBs with pulses of 
longer duration and softer gamma rays.
Observationally the only energies of GRBs with long duration  
have been measured. We predict that the energies of GRBs with short duration
are much larger than those of GRBs with longer duration.

We now wish to estimate an energy release rate per unit volume 
and per unit time; the energy released
as ultra high energy cosmic rays.
In order to do so we assume that the dark matter is composed mainly of 
axion stars.
The estimation, however, involves several ambiguities associated with
number density of neutron stars, energy density of dark matter 
or velocity of ASs
in the Universe etc. Therefore the estimation does not lead to 
a conclusive result although our result is consistent with
the observations\cite{uhe}.

First we note that luminosity density 
around our galaxy is observationally given  
by $2\times 10^8\, \mbox{h} \,L_{\odot}/\mbox{Mpc}^3$ 
where h is Hubble constant
with the unit of $100$ km/s Mpc and $L_{\odot}$ represents solar luminosity.
 We take a mass density $\rho$ corresponding 
to this luminosity density such as 
$\rho\sim 10^9M_{\odot}\mbox{h}^2/\mbox{Mpc}^3$. 
Then, we suppose that the number density of 
neutron stars is given by $\sim 10^{-2}\rho$. This comes from the fact that
the rate of appearance of supernovae is about $0.1\sim 1$ 
per $10$ years in our galaxy
and the age of the galaxy is about $10^{10}$ years. 
Probably, the rate of the appearance could be 
larger in early stage of our galaxy than the one at present.
Thus, in our galaxy
$10^9$ neutron stars might be involved corresponding to the 
number $10^{11}$ of stars in the galaxy. 
Hence we guess that the number of the 
neutron stars are equal to about $10^{-2}$ times the number of the stars.
All of these neutron stars are assumed to possess 
strong magnetic field $>10^{12}$ G.
Furthermore,
to estimate the rate of the energy release 
we need to know the average density $\epsilon$ 
of the dark matter.
Here we use a value\cite{text} of $0.5\times 10^{-24}\mbox{g}/\mbox{cm}^3$,
which represents a local density of our halo. 
Using these parameters we can estimate the rate owing to 
the collision between the AS 
and the neutron star if the cross section of
the collision is found. 
The collision cross section for 
a neutron star to trap an AS is estimated 
in the following. Namely an AS is trapped by the neutron star when
the AS approaches the neutrons star within a distance $L_c$ in which
its kinetic energy $M_av^2/2$ is equal to its potential energy 
$\,1.5\times M_{\odot}\,M_a\,G/L_c$ around the neutron star.
$G$ is gravitational constant. Here 
the mass of the neutron star and the relative velocity $v$ are assumed to be 
$1.5\times M_{\odot}$ and $3\times 10^7$ cm/s, respectively. 
Thus, the cross section is found such as $L_c^2\pi$ 
with $L_c\simeq 6\times 10^{11}\,\mbox{cm}$.
After being trapped, the AS collides with
the neutron star by losing its potential energy and angular momentum
owing to the emission of gravitational and electromagnetic waves;
electromagnetic radiations arise due to the oscillating current $J_a$.
The merger is similar to neutron star - neutron star merger\cite{totani}.
Time scale from birth to merger is much less than the age of the Universe
when the distance $L_c$ between two stars at the birth is given such that 
$L_c\sim 10^{11}$ cm.

Therefore,
we obtain the rate of the collision per Mpc$^{3}$ and per year,

\begin{equation}
\epsilon\times 10^{-2}\rho\times L_c^2\pi\times v\times 1\, 
\mbox{year}/10^{-1}M_{\odot}\simeq 3\times 10^{-9}\,\mbox{h}^2/\mbox{Mpc}^3\,
\, \mbox{year}.
\end{equation}
Since the energy of $\sim 10^{53}$ erg is released in the collision, 
we find that the rate of the energy release is given by 

\begin{equation}
\sim3\times 10^{44}\,\mbox{h}^2\, \mbox{erg}/\mbox{Mpc}^3\,\,\mbox{year}.
\end{equation} 
which agrees well with the observed one . 
Taking account of the fact, however, that there are several ambiguities in the 
parameters used above and in the observations, we understand that
our model can explains roughly 
the observations\cite{uhe}. 
%For instance,
%if the velocity $v$ is smaller by factor $3$ than the above one, the rate
%increases $3$ times.

Finally we discuss two possibilities of the observation of the axion stars.
One is associated with gravitational lensing and the other one associated
with monochromatic radiations from the ASs.
 
Since we assume that the halo of our galaxy is composed of ASs and
their mass is $\sim 10^{-1}M_{\odot}$, 
the ASs are plausible candidates 
for gravitational microlenses, i.e. MACHO.
Since baryonic candidates like white dwarfs, neutron stars, etc. have serious 
problems, nonbaryonic candidates are favored\cite{non}. 
The problems are associated with 
chemical abundance of carbon and nitrogen in the Universe: If these baryonic 
stars are MACHOs, 
an overabundance of the elements is produced far in excess of what
is observed in our galaxy.
%With the choice of the axion mass, $10^{-9}$ eV, the ASs become possible 
%candidates since their mass is approximately given by $10^{-1}M_{\odot}$. 
Hence, the ASs are theoretically the most fascinating candidates for
MACHOs since they are also candidates for the generators of 
both UHECRs and GRBs.
If the most favorable mass of the MACHO is  
$0.5M_{\odot}$,
we need to choose 
$m_a\simeq 0.2\times 10^{-9}$ eV 
since the mass of the ASs is given by $\simeq 10^{-1}M_{\odot}/m_9$. 
Smaller axion mass leads to stronger electric field as 
well as larger mass of AS. Thus,
it yields 
higher energies of the cosmic rays, $\sim 5\times 10^{20}$ eV 
and of GBRs, $\sim 5\times 10^{53}$ erg than the ones 
we have claimed above. Accordingly, 
the determination of the mass of MACHOs 
gives the upper limit of both the energies of the ultra high energy 
cosmic rays and the energies released in the gamma ray bursts in our
mechanism.

We point out another way of the observation of the axion stars. Since 
the electric field induced in ASs is oscillating, 
electromagnetic radiations
are emitted\cite{iwasol,rad} from corresponding oscillating electric current 
$J_a$ as mentioned above; 
the frequence of these monochromatic radiations is 
$\simeq 2.4\times 10^5\,m_9$ Hz. 
We expect that such radiations can be detected 
in advance of UHECRs; they are emitted by the AS revolving 
around a neutron star
before the AS decaying into charged particles. It is easy to 
estimate their luminocity\cite{rad} 
$\simeq 6.7\times 10^{41}\,B_{12}^2\,\mbox{erg/s}$. If the emission arises 
in a distance $\sim 10$ Mpc from the earth, we obtain the flux at the earth,
$\sim 10^9\,\mbox{Jy}\,B_{12}^2/m_9$; we have assumed that 
the velocity of the AS revolving 
is $\sim 0.1 \times $ light velocity. 
Although the possibility of observing the radiations 
is very intriguing,
it might be difficult to
detect the radiations with such a low 
frequency because they could be absorbed by interstellar 
ionized gases.

\vskip .5cm

The author wishs to express his thank to all the staffs in Theory Group,
Tanashi Branch, High Energy Accelerator Research Organization 
for their warm hospitality extended to him.
This work is supported by the Grant-in-Aid for Scientific Research
from the Ministry of Education, Science and Culture of Japan No.10640284

%%%%%%%%%%%%%%%%%%%%%%


\begin{thebibliography}{99}
\bibitem{uhe}
M.A. Lawrence et al. J. Phys. G. Nucl. Part. Phys. 17, 773 (1991),\\
D.J. Bird et al. Phys. Rev. Lett. 71, 3401 (1993); ApJ 424, 491 (1994),\\
N. Hayashida et al. Phys. Rev. Lett. 73, 3491 (1994),\\
M. Takeda et al. Phys. Rev. Lett. 81, 1163 (1998); ApJ, 522, 225 (1999).
%\bibitem{model}P. M\'esz\'aros, astro-ph/9904038, and references cited therein.
\bibitem{text}For a review, see, for example, 
E.W. Kolb and M.S. Turner, The Early Universe, Addison-Wesley, New York,
(1990).
\bibitem{PQ}R.D. Peccei and H.R. Quinn, Phys. Rev. Lett. 38, 1440 (1977),\\
S. Weinberg, Phys. Rev. Lett. 40, 223 (1978),\\
F. Wilczeck, Phys. Rev. Lett. 40, 279 (1978).
\bibitem{kim}J.E. Kim, Phys. Rep. 150, 1 (1987).
\bibitem{cooling}E. Seidel and  W.M. Suen, Phys. Rev. Lett. 72, 2516 (1994).
\bibitem{kolb}E.W. Kolb and I.I. Tkachev, Phys. Rev. Lett. 71, 3051 (1993);\\
Phys. Rev. D49, 5040 (1994).
\bibitem{iwa}A. Iwazaki, Phys. Lett. B455, 192 (1999).
\bibitem{iwa1}A. Iwazaki, Phys. Rev. D60 025001 (1999).
\bibitem{string}T. Banks and M. Dine, Nucl. Phys. B479, 173 (1996);
Nucl. Phys. B505, 445 (1997).
\bibitem{MACHO}C. Alcock et al. Astrophys. J. 449, 28 (1995),\\
for a review, see  C.S. Kochanek and J.N. Hewitt, eds, ``Astrophysical 
Applications of Gravitational Lensing'', IAU Symp. 173 (1996) 
(Kluwer, Dordrecht).
\bibitem{non}see, for example, F. Katherine astro-ph/9901178.
\bibitem{iw}A. Iwazaki, Prog. Theor. Phys. 101, 1253 (1999). 
\bibitem{vacdecay}J. Schwinger, Phys. Rev. 82, 664 (1951).
\bibitem{real}E. Seidel and W.M. Suen, Phys. Rev. Lett. 66, 1659 (1991).
\bibitem{iwasol}A. Iwazaki, Phys. Lett. B451, 123 (1999).
\bibitem{entropy}P.J. Steinhardt and M.S. Turner, Phys. Lett. B129 51, (1983).
%\bibitem{complex}B. Stern, astro-ph/9902203.
\bibitem{re}P. Jetzer, Phys. Rep. 220, 163 (1992),\\ T.D. Lee 
and Y. Pang, Phys. Rep.
221, 251 (1992),\\ A.R. Liddle and M. Madsen, 
Int. J. Mod. Phys. D1, 101 (1992).
\bibitem{star}S.L. Shapiro and S.A. Teukolsky, Black Holes, White Dwarfs, 
and Neutron Stars, A Wiley-Interscience Publication, (1983).
\bibitem{DFSZ}A.R. Zhitnitsky, Sov. J. Nucl. Phys. 31, 260 (1980),\\ M.D. Dine,
W. Fischler and M. Srednicki, Phys. Lett. B104, 199 (1981).
\bibitem{hadron}J. E. Kim, Phys. Rev. Lett. 43, 103 (1979),\\ M.A. Shifman,
A.I. Vainshtein and V.I. Zakharov, Nucl. Phys. B166, 493 (1980).
\bibitem{Si}P. Sikivie, Phys. Lett. B137, 353 (1984). 
%\bibitem{s}P. Sikivie, Phys. Rev. Lett. 51, 1415 (1983).
%\bibitem{decay}I.I. Tkachev, Phys. Lett. B191, 41 (1987).
\bibitem{con}N. Itho et al. ApJ 418, 405 (1993),\\
D.A. Baiko and D.G. Yakovlev, Astron. Lett. 72, 702 (1995);
astro-ph/9604164.
\bibitem{totani}T, Totani, ApJ. 511, 41 (1999).
\bibitem{rad}A. Iwazaki, hep-ph/9908468. 
%\bibitem{ba}M.J. Ree and P. Meszaros, MNRAS, 41P, 258 (1992).
%\bibitem{grb}S. Odewahn, J. Bloom and S. Kulkarni, G.C.N. 201 (1999).
%\bibitem{kolb2}E.W. Kolb and I.I. Tkachev, Phys. Rev. D50, 769 (1994).



\end{thebibliography}
\end{document}